\documentstyle[12pt]{article}
\newlength{\extraspace}
\newlength{\extraspaces}
\newcommand{\ba}{\begin{eqnarray}
\addtolength{\abovedisplayskip}{\extraspaces}
\addtolength{\belowdisplayskip}{\extraspaces}
\addtolength{\abovedisplayshortskip}{\extraspace}
\addtolength{\belowdisplayshortskip}{\extraspace}}
\newcommand{\ea}{\end{eqnarray}}

\newcommand{\nonu}{\nonumber \\[.5mm]}
\newcommand{\A}{&\!\!\!}

\begin{document}

\thispagestyle{empty}

\hfill \parbox{3.5cm}{hep-th/0012235 \\ SIT-LP-00/12}
\vspace*{1cm}
\begin{center}
{\bf  ON GRAVITATIONAL INTERACTION OF     \\ 
SPIN 3/2 NAMBU-GOLDSTONE FERMION } \\[20mm]
{Kazunari SHIMA and Motomu TSUDA} \\[2mm]
{\em Laboratory of Physics,  Saitama Institute of Technology}\footnote{e-mail: shima@sit.ac.jp, tsuda@sit.ac.jp}\\
{\em Okabe-machi, Saitama 369-0293, Japan}\\[2mm]
{December  2000}\\[15mm]

%\maketitle

\begin{abstract}
A new gravitational interaction of spin 3/2  Nambu-Goldstone(N-G) fermion 
is constructed, which gives a new framework for the consistent gravitational 
coupling of spin 3/2 massless field. 
The action is invariant under a new global supersymmetry.        \\

PACS:12.60.Jv, 12.60.Rc, 12.10.-g /Keywords: supersymmetry, Nambu-Goldstone fermion, composite unified theory 
\end{abstract}
\end{center}

\newpage
The supersymmetry(SUSY)[1][2][3] is an essential notion to unify spacetime and matter.
However it is totally unrealistic symmetry  so far in the observed low energy particle physics 
and should be broken spontaneously. 
It is well understood that the Nambu-Goldstone(N-G) fermion with spin 1/2  
would appear in the spontaneous breakdown of SUSY and that it can be converted 
to the longitudunal components of the spin 3/2 field(gravitino) through the superHiggs mechanism. 
This is demonstrated explicitly by the introduction of the local gauge coupling of Volkov-Akulov(V-A) model[2] 
of a nonlinear realization of SUSY(NL SUSY)to the supergravity(SUGRA) gauge multiplet[4]. 

In ref.[5] and [6], we have proposed a supersymmetric composite unified model for spacetime and matter,
superon-graviton model(SGM) based upon SO(10) super-Poincar\'e algebra, 
where we have regarded the  spin 1/2 N-G fermions of V-A NL SUSY[2] 
as the fundamental objects(superon-quintet) for matter. 
SGM may be the most economical model that accomodates all observed particles in a single irreducible 
representation of a (semi)simple group. 
The NL SUSY  may give a framework to describe the unity of nature from the compositeness viewpoint for matter.  
In SGM all particles participating in (super)Higgs  mechanism except graviton are  composites of N-G fermions, superons. 
In ref.[6], we have constructed the gauge invariant SGM action and clarified the systematics in the 
unified model building.

In thie letter we extend the framework[6] to N-G fermion with the higher spin.
Following the arguments of V-A, the action of N-G fermion $\psi^{\mu}_{\alpha}(x)$ with spin 3/2 
is already written down by Baaklini as a nonlinear realization of a new superalgebra containing 
a vector-spinor generator $Q^{\mu}_{\alpha}$[7].
We study in detail the gravitational interaction of Baaklini model[7].
We will see that the similar  arguments to  SGM can be performed and produce a new  gauge invariant action, 
which is the straightforward generalization of SGM action.  The phenomenological implications of spin 3/2 
fundamental constituents are discussed briefly. 

In ref.[7], a new SUSY algebra containing a spinor-vector generator $Q^{\mu}_{\alpha}$ is introduced as follows:
\begin{equation}
%\eqalign{
\{Q^{\mu}_{\alpha},Q^{\nu}_{\beta}\}=\varepsilon^{\mu\nu\lambda\rho}P_{\lambda}(\gamma_{\rho}\gamma_{5}C)_{\alpha\beta}, 
%}
\end{equation} 
\begin{equation}
%\eqalign{
[Q^{\mu}_{\alpha},P^{\nu}]=0,
%}
\end{equation} 
\begin{equation}
%\eqalign{
[Q^{\mu}_{\alpha},J^{\lambda\rho}]={1\over2}(\sigma^{\lambda\rho}Q^{\mu})_{\alpha}+
i\eta^{\lambda\mu}Q^{\rho}_{\alpha}-i\eta^{\rho\mu}Q^{\lambda}_{\alpha},
%}
\end{equation} 
where $Q^{\mu}_{\alpha}$ are vector-spinor generators satisfying Majorana condition 
$Q^{\mu}_{\alpha}=C_{\alpha\beta}{\overline Q}^{\mu}_{\alpha}$, $C$ is a charge conjugation matrix 
and ${1 \over 2}\{\gamma^{\mu},\gamma^{\nu}\}=\eta^{\mu\nu}=(+,-,-,-)$.
By extending  the arguments of V-A model of NL SUSY[2], they obtain the following action 
as the nonlinear representation of the new SUSY algabra. 
\begin{equation}
%\eqalign{
S={1 \over \kappa}\int \omega_{0} \wedge \omega_{1} \wedge \omega_{2} \wedge \omega_{3}
={1 \over \kappa}\int \det{w_{ab}}d^{4}x,
%}
\end{equation} 
\begin{equation}
%\eqalign{
w_{ab}={\delta}_{ab}+ t_{ab},  \quad
t_{ab}=ia \varepsilon_{acde}\bar{\psi}^{c}\gamma^{d}\gamma_{5}\partial_{b}{\psi}^{e},
%}
\end{equation} 
where $\kappa$ and $a$ are  up to now  arbitrary constants with the dimension of the fourth power of length(i.e., 
a fundamental volume of  spacetime) 
and  $\omega_{a}$ is the following differential forms
\begin{equation}
%\eqalign{
\omega_{a} =dx_{a} + ia \varepsilon_{abcd}\bar{\psi}^{b}\gamma^{c}\gamma_{5}d{\psi}^{d},
%}
\end{equation} 
which is invariant under the following (super)translations 
\begin{equation}
%\eqalign{
\psi^{a}_{\alpha} \longrightarrow   \psi^{a}_{\alpha} + \zeta^{a}_{\alpha}, 
%}
\end{equation} 
\begin{equation}
%\eqalign{
x_{a}  \longrightarrow   x_{a} + ia \varepsilon_{abcd}\bar{\psi}^{b}\gamma^{c}\gamma_{5}{\zeta}^{d}, 
%}
\end{equation} 
where $\zeta^{a}_{\alpha}$ is a constant Majorana tensor-spinor parameter.

Now we consider the  gravitational interaction of Baaklini model(4). 
We show that the arguments performed in  SGM[6] of spin 1/2  N-G field $\psi_{\alpha}(x)$ can be extended 
straightforwardly to spin 3/2 Majorana N-G field $\psi^{a}_{\alpha}$. 
In the present case, as seen in (7) and  (8) NL SUSY SL(2C) degrees of freedom 
(i.e. the coset space coordinates  $\psi^{a}_{\alpha}$ representing N-G fermions) in addition to 
Lorentz SO(3,1) coordinates are embedded at every curved spacetime point with GL(4R) invariance.    \\
Following the arguments of SGM[6], it is natural to introduce formally a new vierbein field ${w^{a}}_{\mu}(x)$ through 
the NL SUSY invariant differential forms 
$\omega_{a}$ in (6)  as follows: 
\begin{equation}
%\eqalign{
\omega^{a} = {w^{a}}_{\mu}dx^{\mu}, 
%}
\end{equation} 
\begin{equation}
%\eqalign{
{w^{a}}_{\mu}(x) = {e^{a}}_{\mu}(x) +  {t^{a}}_{\mu}(x),    \quad  
{t^{a}}_{\mu}(x)=ia \varepsilon^{abcd}\bar{\psi}_{b}\gamma_{c}\gamma_{5}\partial_{\mu}{\psi}_{d}, 
%}
\end{equation} 
where ${e^{a}}_{\mu}(x)$ is the vierbein of Einstein Genaral Relativity Theory(EGRT) and 
Latin $(a,b,..)$ and Greek $(\mu,\nu,..)$ are the indices for local Lorentz and general coordinates, respectively.
By noting $(\psi^{\mu}_{\alpha}(x))^{2}=0$, we can easily obtain the inverse of  the new vierbein, ${w_{a}}^{\mu}(x)$, 
in the power series of ${t^{a}}_{\mu}$ which terminates with $({t})^{4}$: 
\begin{equation}
%\eqalign{
{w_{a}}^{\mu} = {e_{a}}^{\mu} - {t^{\mu}}_{a} +{t^{\rho}}_{a}{t^{\mu}}_{\rho}- \dots . 
%}
\end{equation}  
Similarly we introduce formally a new metric tensor  $s^{\mu\nu}(x)$ in the abovementioned curved spacetime 
as follows:
\begin{equation}
%\eqalign{
s^{\mu\nu}(x) \equiv {w_{a}}^{\mu}(x) w^{{a}{\nu}}(x). 
%}
\end{equation}
It is easy to show 
${w_{a}}^{\mu} w_{{b}{\mu}} = \eta_{ab}$,  $s_{\mu \nu}{w_{a}}^{\mu} {w_{b}}^{\mu}= \eta_{ab}$, ..etc. 
In order to obtain simply the action in the abovementioned curved spacetime, which is invariant  at least 
under GL(4R), NL SUSY and local Lorentz transformations, we follow formally EGRT as performed in SGM[6]. 
That is, we require that the (mimic) vierbein ${w^{a}}_{\mu}(x)$ and the  metric  
$s^{\mu\nu}(x)$ should have formally a general coordinate transformations  under the supertranslations: 
\begin{equation}
%\eqalign{
\delta x_{\mu}= - \xi_{\mu}, \quad 
\delta\psi^{a}=\zeta^{a},
%}
\end{equation}wher
where $\xi^{\mu}=-ia \varepsilon^{\mu\nu\rho\sigma}\bar{\psi}_{\nu}\gamma_{\rho}\gamma_{5}{\zeta}_{\sigma}$.  \\
Remarkably we find that the following nonlinear transformations 
\begin{equation}
%\eqalign{
\delta \psi^{a}(x) = \zeta^{a} - ia (\varepsilon^{\mu\nu\rho\sigma}\bar{\psi}_{\nu}\gamma_{\rho}
\gamma_{5}{\zeta}_{\sigma})\partial_{\mu}\psi^{a}
%}
\end{equation} 
\begin{equation}
%\eqalign{
\delta {e^{a}}_{\mu}(x) = ia (\varepsilon^{\rho\nu\sigma\lambda}\bar{\psi}_{\nu}\gamma_{\sigma}
\gamma_{5}{\zeta}_{\lambda})D_{[\mu}{e^{a}}_{\rho]}
%}
\end{equation} 
induce the desirable transformations on ${w^{a}}_{\mu}(x)$ and $s^{\mu\nu}(x)$ as follows: 
\begin{equation}
%\eqalign{
\delta_{\zeta_{1}} {w^{a}}_{\mu} = \xi^{\nu}_{1} \partial_{\nu}{w^{a}}_{\mu} + \partial_{\mu} \xi^{\nu}_{1} {w^{a}}_{\nu}, 
%}
\end{equation} 
\begin{equation}
%\eqalign{
\delta_{\zeta_{1}} s_{\mu\nu} = \xi^{\kappa}_{1} \partial_{\kappa}s_{\mu\nu} +  
\partial_{\mu} \xi^{\kappa}_{1} s_{\kappa\nu} 
+ \partial_{\nu} \xi^{\kappa}_{1} s_{\mu\kappa}, 
%}
\end{equation} 
where also throughout the paper $D_{\mu}$ is the covariant derivative of GL(4,R) with the symmetric affine connection 
and $\xi^{\rho}_{\zeta_{1}}=-ia \varepsilon^{\mu\nu\rho\sigma}\bar{\psi}_{\nu}\gamma_{\rho}\gamma_{5}{\zeta}_{1\sigma}$. 
That is, ${w^{a}}_{\mu}(x)$ and $s^{\mu\nu}(x)$  have general coordinate transformations under 
the new supertransformations (14) and (15).                    \\

Therefore replacing  ${e^{a}}_{\mu}(x)$ in Einstein-Hilbert Lagrangian of general relativity 
by  the new vierbein ${w^{a}}_{\mu}(x)$  
we obtain the following Lagrangian which is invariant under (14) and (15):

\begin{equation}
L=-{c^{3} \over 16{\pi}G}\vert w \vert(\Omega + \Lambda ),
\end{equation}
\begin{equation}
%\eqalign{
\vert w \vert=det{w^{a}}_{\mu}=det({e^{a}}_{\mu}+{t^{a}}_{\mu}),  
%}
\end{equation} 
where the overall factor is now fixed uniquely to ${-c^{3} \over 16{\pi}G}$, 
${e_{a}}^{\mu}(x)$ is the vierbein of EGRT and 
$\Lambda$ is a probable cosmological constant. 
$\Omega$ is a mimic new scalar curvature analogous to the Ricci scalar curvature $R$ of EGRT. The explicit 
expression of $\Omega$ is obtained  by just replacing ${e_{a}}^{\mu}(x)$ in Ricci scalar $R$ of EGRT by 
${w_{a}}^{\mu}(x)={e^{a}}_{\mu}+{t^{a}}_{\mu}$, which gives the gravitational interaction of  $\psi^{a}_{\alpha}(x)$. 
The lowest order term of $a$ in the action (11) gives the Einstein-Hilbert 
action of general relativity. And in flat spacetime, i.e.  ${e_{a}}^{\mu}(x) \rightarrow {\delta_{a}}^{\mu}$, 
the action (11) reduces to  V-A model[2] with  ${\kappa}^{-1} = {c^{3} \over 16{\pi}G}{\Lambda}$. 
Therefore our model predicts  a non-zero (small) cosmological constant.                                       \\
The commutators of two new supersymmetry transformations  on $\psi^{a}_{\alpha}(x)$ and  ${e_{a}}^{\mu}(x)$ 
are 

\begin{equation}
%\eqalign{
[\delta_{\zeta_{1}}, \delta_{\zeta_{2}}] \psi^{a} = 
\{2ia (\varepsilon^{{\mu}bcd}\bar{\zeta}_{2b}\gamma_{c}\gamma_{5}{\zeta}_{1d})
- \xi_{1}^{\rho} \xi_{2}^{\sigma} {e_{a}}^{\mu} (D_{[\rho} {e^{a}}_{\sigma]})\} \partial_{\mu} \psi^{a}, 
%}
\end{equation} 

\begin{equation}
%\eqalign{
%}
[\delta_{\zeta_{1}}, \delta_{\zeta_{2}}] {e^{a}}_{\mu} = 
\{2ia (\varepsilon^{{\rho}bcd}\bar{\zeta}_{2b}\gamma_{c}\gamma_{5}{\zeta}_{1d})
- \xi_{1}^{\sigma} \xi_{2}^{\lambda} {e_{c}}^{\rho} (D_{[\sigma} {e^{c}}_{\lambda]})\} 
D_{[\rho} {e^{a}}_{\mu]}
 -  \partial_{\mu}(  \xi_{1}^{\rho} \xi_{2}^{\sigma} D_{[\rho} {e^{a}}_{\sigma]}). 
%}
\end{equation} 
The equations (14),(15),(20) and (21) may reveal N-G fermion (NL SUSY) nature of $\psi^{a}_{\alpha}(x)$, 
non-N-G  nature of ${e_{a}}^{\mu}(x)$  and a generalized general coordinate- 
and local Lorentz-transformations, which form a closed algebra[8].          \\
Although the discussions are  displayed in parallel with spin 1/2 SGM case[6] for simplicity 
and generality for N-G fields, it is remarkable that the  ${\em massless}$  spin 3/2 
field realized as a N-G fermion of NL SUSY  can have a new consistent gravitational coupling 
without using SUGRA framework in which the local gauge invariance is respected 
for ${\em massless}$ spin 3/2 gravitino field.      \\ 
It is interesting to expand tentatively SGM action explicitly in terms of  
${e^{a}}_{\mu}(x)$ and ${t^{a}}_{\mu}(x)$ in order to clarify the differences of 
the gravitational interaction of ${\em massless}$ spin 3/2 field in between SGM and SUGRA.  
We obtain      \\   
\ba
%\eqalign{
%}
L_{SGM} = \A \A - {c^3\Lambda \over 16{\pi}G} e \vert w \vert - {c^3 \over 16{\pi}G} e R 
\nonu
\A \A - {c^3 \over 16{\pi}G} e \{ t^{\mu\nu}R_{\mu\nu} +  t^{\mu\rho}{t^{\nu}}_{\rho}R_{\mu\nu} + O(t^{2}) 
+ O(t^{3}) +  \dots  + O(t^{16}) \},
%}
\ea
where  $\vert w \vert = det{w^{a}}_{\mu}$, $e=det{e^{a}}_{\mu}$ and $R$ and $R_{\mu\nu}$ 
are the Ricci curvature tensors  of GR. 
The first term reduces to Baaklini action[6] with ${\kappa}^{-1} = {c^3 \over 16{\pi}G}{\Lambda}$ 
in the Riemann-flat $e{_a}^{\mu}(x) \rightarrow \delta{_a}^{\mu}$ spacetime and  
the second term is the familiar E-H action of GR. 
We can easily see the  complementary relation of graviton and superon(in the form of the stress energy tensor) 
in SGM.  
These situations are apparently rather different from SUGRA and remain to be studied in detail.     \\
Finally we just mention the phenomenological implications of our model. 
As read off from the above discussions it is easy to introduce (global) SO(N) internal symmeytry in our model 
by replacing $\psi^{a}_{\alpha}(x) \rightarrow  {\psi^{ia}}_{\alpha}(x),(i=1,2, \dots, N)$,  
which may enable us to consider  SGM[5] with spin 3/2 superon. However the fundamental internal symmetry 
for superons may be rather different from spin 1/2 SGM. For the generator of a new algebra shifts spin by 3/2 
and one-superon states correspond to spin 1/2 states, although the adjoint multiplet ${N(N-1) \over 2}$ 
appear at the vector state.      \\ 
Also apart from the composite SGM scenario, it is worthwhile 
to consider SGM with  extra spacetime dimensions and its compactifications.

\vskip 30mm

The work of M. Tsuda is supported in part by  High-Tech research program of
Saitama Institute of Technology.

\newpage

%%%%%%%  References  %%%%%%%%%%%%%%%%%%%%%%%%%%%%%%%%%%%%%%%
%
\newcommand{\NP}[1]{{\it Nucl.\ Phys.\ }{\bf #1}}
\newcommand{\PL}[1]{{\it Phys.\ Lett.\ }{\bf #1}}
\newcommand{\CMP}[1]{{\it Commun.\ Math.\ Phys.\ }{\bf #1}}
\newcommand{\MPL}[1]{{\it Mod.\ Phys.\ Lett.\ }{\bf #1}}
\newcommand{\IJMP}[1]{{\it Int.\ J. Mod.\ Phys.\ }{\bf #1}}
\newcommand{\PR}[1]{{\it Phys.\ Rev.\ }{\bf #1}}
\newcommand{\PRL}[1]{{\it Phys.\ Rev.\ Lett.\ }{\bf #1}}
\newcommand{\PTP}[1]{{\it Prog.\ Theor.\ Phys.\ }{\bf #1}}
\newcommand{\PTPS}[1]{{\it Prog.\ Theor.\ Phys.\ Suppl.\ }{\bf #1}}
\newcommand{\AP}[1]{{\it Ann.\ Phys.\ }{\bf #1}}

\end{document}